\documentclass{elphreport_2019}
\usepackage{graphicx} 

\begin{document}

\pacs{2935} 

\title{Development of novel silicon sensors with high time and spatial resolution}


\author[label1]{Taikan Suehara}
\author[label2]{Yuto Deguchi}
\author[label2]{Yuto Uesugi}
\author[label3]{Yu Kato}
\author[label4]{Ryo Yonamine}

\address[label1]{Department of Physics, Faculty of Science, Kyushu University, Fukuoka, 819-0395}
\address[label2]{Department of Physics, Graduate School of Science, Kyushu University, Fukuoka, 819-0395}
\address[label3]{Department of Physics, Graduate School of Science, The University of Tokyo, Tokyo, 113-0033}
\address[label4]{Department of Physics, Graduate School of Science, Tohoku University, Sendai, 980-8578}

\maketitle

\begin{abstract}
 Silicon pad sensors with novel functions of higher timing resolution (LGAD: Low Gain Avalanche Detector) and higher position resolution (PSD: Position Sensitive Detector)
 are studied for an application to Silicon-Tungsten electromagnetic calorimeter for a detector of the International Linear Collider (ILC).
 Prototype sensors are fabricated, equipped with dedicated ASICs (Application-Specific Integrated Circuits) and tested with a positron beam as well as a radioisotope.
 The first results of the measurements of timing resolution with LGADs and position reconstruction with PSDs are reported.
\end{abstract}

\section{Introduction}

The International Linear Collider (ILC) is a future electron-positron collider for precise measurements of Higgs bosons and various BSM searches. Silicon-tungsten electromagnetic calorimeter (SiW-ECAL) is one of the candidates to be used in the International Large Detector (ILD)\cite{a}, one of the detector concepts for the ILC. 
We are proposing two types of new silicon sensors, Low Gain Avalanche Detectors (LGADs) and Position Sensitive Detectors (PSDs) to be replaced to the standard silicon pads of SiW-ECAL. LGADs are possible to obtain precise timing information and PSDs are sensitive to hit positions within a cell. With LGADs and PSDs position and timing resolution of the SiW-ECAL can be significantly improved, which should proceed further reach of Higgs and BSMs in the ILC physics program.

LGADs are silicon sensors with the internal avalanche amplification, which have already been proved to realize the timing resolution down to 30 psec\cite{lgad}.
The precise timing information can be primarily utilized to perform particle identification of hadrons by Time-of-Flight (ToF) method, which can be
combined with $dE/dx$ information obtained in the Time Projection Chamber, which is the main tracking detector in the ILD.

PSDs are silicon sensors with each cell having an electrode at each corner instead of a simple pad spread over the cell. When the signal charge reaches P$^+$ pad, the charge is resistively split to electrodes via a resistive layer on the surface. The hit position is reconstructed as the gravity center of signal strengths of the electrodes at the four corners. 
In contrast to using smaller cells, the position resolution can be improved with minimal increase of the readout channels if we replace the silicon pads with PSDs in SiW-ECAL.

We prepared samples for both LGADs and PSDs to demonstrate the possibilities to be used for SiW-ECAL and to measure
characteristics of the sensors for the optimization. We have conducted measurements with the positron beam provided from ELPH
as well as particles from radioisotopes.

\section{Setup of the test beam}

 \begin{figure}[!ht]
  \begin{center}
   \includegraphics[width=0.4\textwidth]{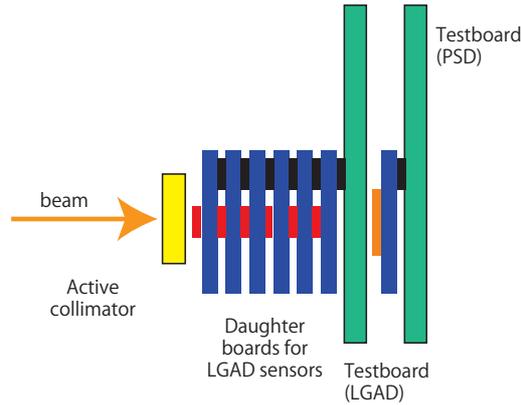}
   \caption{Overview of the setup of LGAD/PSD measurement. The positron beam penetrates up to 6 LGADs and a PSD to obtain concurrent signal.}
   \label{FIG:overview}
  \end{center}
 \end{figure}

Figure \ref{FIG:overview} shows the overview of the setup.
The LGADs and PSDs are equipped with testboards of Skiroc2-CMS ASIC (Application-Specific Integrated Circuit).
Skiroc2-CMS has 64-channel readout for silicon sensors with precise charge and timing information.
The timing jitter is $\sim$30 psec with charge of $>$ 150 fC with dynamic range of ADC $>$ 1000 fC.
The timing information is obtained via clock count of 40 MHz and analog 12-bit Time-of-Arrival (ToA),
which clips the voltage sweeping within a clock. The non-linearity of ToA and time walk are corrected
 using measurements with charge injection.

Up to 6 LGADs were stacked with single testboard to obtain correlation of the signal from penetrating beam.
The testboards with LGADs and PSDs were synchronized with a common-stop signal.
Self-triggering with LGADs were used for the common stop.
An active collimator, which is a plastic scintillator with a center hole, placed in front of the sensor stack.
The optical readout of the collimator is done by PMT, connected to a remaining channel of the testboard.

\section{Results with the LGADs}

 \begin{table}[!ht]
  \caption{List of LGADs and APDs tested with beam at ELPH, Tohoku University.}
  \label{TBL:sensors}
  \begin{center}
   \begin{tabular}{c|c|c|r|r|r|r}\hline
    Spec No. & Type & Size & Count \#1 & Count \#2 & Count \#1\&\#2 & Efficiency\\ \hline\hline
    S12023-10A & Reach-through & $\phi$1 & 1002 & 965 & 147 & 14.9\% \\ \hline
    S2384               & Reach-through & $\phi$3 & 4355 & 5796 & 1136 & 22.4\% \\ \hline
    S8664-10K  & Inverse & $\phi$1 & 613 & 298 & 4 & 0.9\% \\ \hline
    S8664-20K  & Inverse & $\phi$2 & 368 & 185 & 2 & 0.7\% \\ \hline
    S8664-55  & Inverse & 5$\times$5 & 3060 & 2327 & 96 & 3.6\% \\ \hline
    pkg-10 & Reach-through & $\phi$1 & 1687 & 1584 & 15 & 0.9\% \\ \hline
    pkg-20 & Reach-through & $\phi$1 & 1956 & 3010 & 219 & 8.8\% \\ \hline
   \end{tabular}
  \end{center}
 \end{table}

The sensors we have tested at the beam is listed in Table \ref{TBL:sensors}\cite{chef_lgad}.
Some sensors are designed as Avalanche Photo-Diodes (APDs) which are used for optical photon
measurements with avalanche gain. We recognized those as prototype of LGADs since
the basic structure is the same. There are also prototypes of LGADs developed in Hamamatsu
(pkg-10 and pkg-20) which were also tested with beam.
There are two types of LGADs/APDs, reach-through type and inverse type.
The reach-through type has proven performance of timing resolution, but since the amplification region
is confined below the readout pads, there is significant inactive area. In contrast, the inverse type has
amplification on the bottom of the sensor, which should result in more uniform response over the surface.

The Table \ref{TBL:sensors} also shows the number of single and coincidence counts with the beam injection.
The sensors are placed to be penetrated by the beam, but we have no selection of the angle of the beam so
some fraction of the particles can geometrically hit only one sensor, depending on the sensor size.
From the table we see coincidence ratio of the inverse APDs are significantly smaller than the reach-through APDs.
The possible reason is hit-by-hit fractuation of the gain at the inverse-type APDs, but more investigation is
necessary to confirm this.

 \begin{figure}[!ht]
  \begin{center}
   \includegraphics[width=0.9\textwidth]{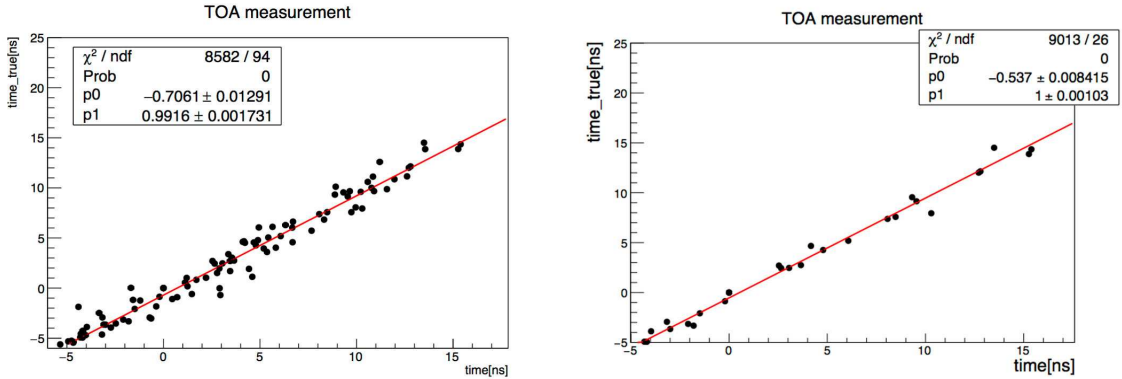}
   \caption{Timing correlation of two APDs (S2384) without selection (left) and with selection of $> 100$ fC (right). }
   \label{FIG:timing}
  \end{center}
 \end{figure}

Timing resolution of reach-through APDs (S2384) were measured with the ToA data of the coincidence events.
After the correction of non-linearity and time-walk of the ToA, we see clear correlation between two sensors
as shown in Figure \ref{FIG:timing} (left). Figure \ref{FIG:timing} (right) shows the correlation of events
having $> 100$ fC of charge to avoid jitter of the electronics. The timing resolution of single sensor was
calculated as $385 \pm 94$ psec. The result is worse than expected. Possible reason is imperfect correction
of ToA since the correction coefficients are obtained with measurements at different places.
We will investigate this with further measurements.

\section{Results with the PSDs}

 \begin{figure}[!ht]
  \begin{center}
   \includegraphics[width=0.5\textwidth]{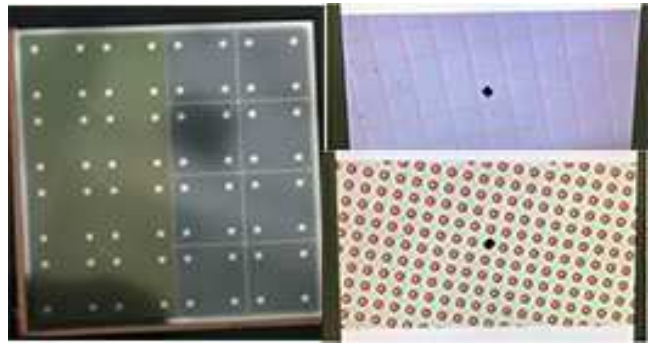}
   \caption{Picture of a PSD (left) and magnified view of resistive P$^+$ surface (top-right, left-half of the PSD) and resistive layer surface (bottom-right, right-half of the PSD). }
   \label{FIG:PSD}
  \end{center}
 \end{figure}

We fabricated a few types of PSDs having 4$\times$4 cells, shown in Figure \ref{FIG:PSD}. Each cell has an electrode at each corner,
having 64 readout channels in total. One of the important parameters of the PSDs is the surface resistance.
The PSDs have two types of the resistive layer in each side. One is a resistive P$^+$ surface, which forms meshes
to increase the resistivity. The other is a dedicated resistive layer connected to a dotted P$^+$ layer, which
gives 10-30 times more resistance\cite{chef_psd}.
The resistivity of the edges of the PSDs is set lower to reduce position distortion.
The cell size is $5.5 \times 5.5$ mm$^2$, and the sensor thickness is 650 $\mu$m for all the sensors.

 \begin{figure}[!ht]
  \begin{center}
   \includegraphics[width=0.8\textwidth]{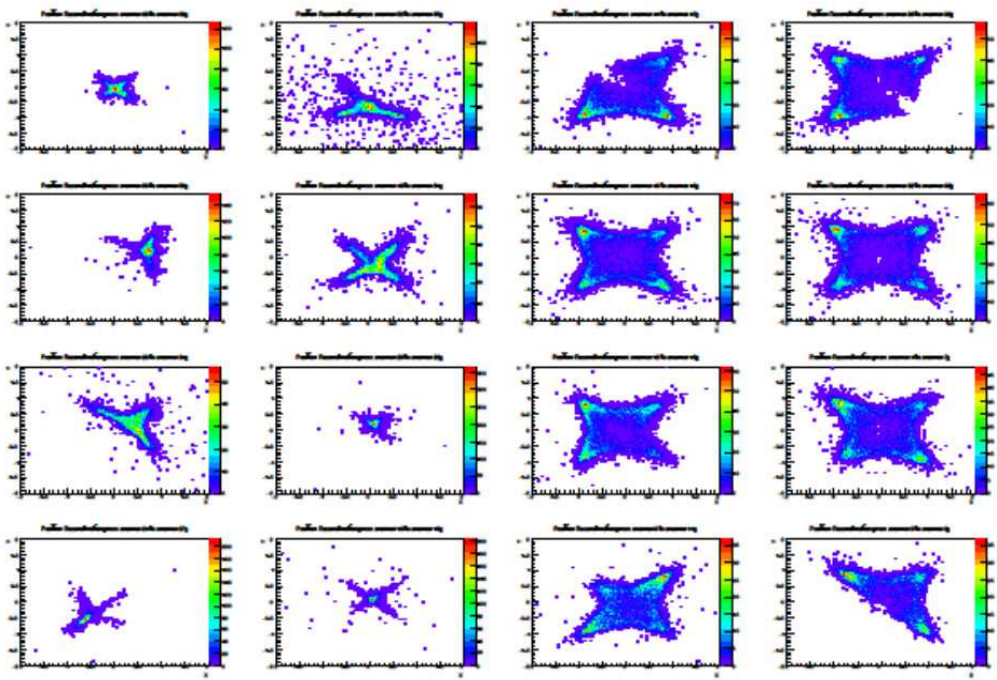}
   \caption{Reconstructed position of each cell with a PSD, with beta radiation from $^{90}$Sr. }
   \label{FIG:PSDdist}
  \end{center}
 \end{figure}

Figure \ref{FIG:PSDdist} shows the response of the PSD sensors with a $^{90}$Sr radioactive source.
The position is calculated by
\begin{eqnarray*}
  X_{\rm{rec}} &=& \rm{\frac{(Q_0+Q_1)-(Q_2+Q_3)}{Q_0+Q_1+Q_2+Q_3}} \\
  Y_{\rm{rec}} &=& \rm{\frac{(Q_0+Q_2)-(Q_1+Q_3)}{Q_0+Q_1+Q_2+Q_3}} 
\end{eqnarray*}
where $X_{\rm{rec}}$ and $Y_{\rm{rec}}$ are the reconstruction position in $X$ and $Y$ axes and $Q_i$ is measured charge at each electrode on the corner.
The difference between maximum and minimum $X_{\rm{rec}}$ ($Y_{\rm{rec}}$), called ``dynamic range" should be $-1$ to $1$ in an ideal case.
The figure shows almost maximum dynamic range is obtained with the resistive layer (right side of the figure).
The concentration of the events at $X_{\rm{rec}} = Y_{\rm{rec}} = \pm 1$ is because of a threshold effect of the trigger on single channel instead of
sum of four channels in a same cell, due to the limitation of the electronics.

The PSDs were also irradiated with positron beam, but due to higher environmental noise at beam line it is difficult to reconstruct positions with
the obtained results. Further studies are expected with more appropriate electronics and shielding.

\section{Summary}

LGADs/APDs and PSDs sensors have been studied for the application to ILD SiW-ECAL.
The first measurement of the timing resolution of the APDs with positron beam is obtained as $385 \pm 94$ psec,
which we will try to improve with updated setup.
We successfully reconstructed the hit positions on PSDs with good dynamic range using a $^{90}$Sr radioactive source.
Further studies with positron beam are planned.

\section*{Acknowledgments}
The test beam experiment was conducted with the support of ELPH, Tohoku University.
We appreciate Omega group for the support on the operation of the Skiroc2-CMS chip.
This work is partially supported by JSPS KAKENHI Grant Number JP17H05407.

\end{document}